\documentclass[11pt]{article}
\usepackage{arxiv}
\usepackage[dvips]{graphicx}
\usepackage{url}
\usepackage{siunitx}
\usepackage{cite}

\begin{document}
%
\title{Analog-to-Stochastic Converter Using Magnetic Tunnel Junction Devices  for Vision Chips}




%

\author{
Naoya Onizawa \\
Frontier Research Institute for Interdisciplinary Sciences, Tohoku University \\
Sendai, Miyagi 980-8578, Japan \\
\texttt{nonizawa@m.tohoku.ac.jp}
\And
Daisaku Katagiri \\
Research Institute of Electrical Communication, Tohoku University \\
Sendai, Miyagi 980-8577, Japan \\
\texttt{katagiri@ngc.riec.tohoku.ac.jp}
\And
Warren J. Gross \\
Department of Electrical and Computer Engineering, McGill University \\
Montreal, Quebec H3A 0E9, Canada \\
\texttt{wjgross@mcgill.ca}
\And
Takahiro Hanyu \\
Research Institute of Electrical Communication, Tohoku University \\
Sendai, Miyagi 980-8577, Japan \\
\texttt{hanyu@ngc.riec.tohoku.ac.jp}
}
\date{}


\maketitle

\begin{abstract}
This paper introduces an analog-to-stochastic converter using a magnetic tunnel junction (MTJ) device for vision chips based on stochastic computation.
Stochastic computation has been recently exploited for area-efficient hardware implementation, such as low-density parity-check (LDPC) decoders and image processors.
However,  power-and-area hungry two-step (analog-to-digital and digital-to-stochastic) converters are required for the analog to stochastic signal conversion.
To realize a one-step conversion, an MTJ device is used as it inherently exhibits a probabilistic switching behavior between two resistance states.
Exploiting the device-based probabilistic behavior, analog signals can be directly and area-efficiently converted to stochastic signals to mitigate the signal-conversion overhead.
The analog-to-stochastic signal conversion is theoretically described and the conversion characteristic is evaluated using device and circuit parameters.
In addition, the resistance variability of the MTJ device is considered in order to compensate the variability effect on the signal conversion.
Based on the theoretical analysis, the analog-to-stochastic converter is designed  in 90nm CMOS and 100nm MTJ technologies and is verified using a SPICE  simulator (NS-SPICE) that handles both transistors and MTJ devices.

\end{abstract}




%
\section{Introduction}

Stochastic computation \cite{stochastic_first,stochastic} has recently been exploited for area-efficient hardware implementation using probabilities represented by a random sequence of bits, called a {\it Bernoulli} sequence.
Some applications of stochastic computation are low-density parity-check (LDPC) decoders \cite{ldpc1,ldpc2,ldpc3,ldpc4,ldpc5} and image processors \cite{image1,image2,gabor}.
Stochastic computation tends to be more error-tolerant than a conventional binary system as a bit-flip in stochastic computation corresponds to a least-significant-bit (LSB) error while a bit-flip on a most-significant-bit (MSB) error heavily affects a value in the conventional system \cite{stoc_fault}.

The hardware based on stochastic computation is efficiently implemented, but an interface circuit to the stochastic hardware is required.
In the stochastic image processors, analog signals from image sensors are converted to digital signals using analog-to-digital converters (ADC).
Then, the digital signals are converted to stochastic bit streams using digital-to-signal converters that tend to be large because they consist of registers, linear-feedback shift registers (LFSRs), and comparators.
To take advantage of the area efficiency on stochastic computation, the signal-conversion overhead needs to be mitigated.
In \cite{memristor-stochastic}, a concept of the analog-to-stochastic conversion has been proposed using memristors.
However,  the switching behavior of the  memristor is very slow (the order of ms).

\begin{figure}[t]
\begin{center}
\includegraphics[width=1.0\linewidth]{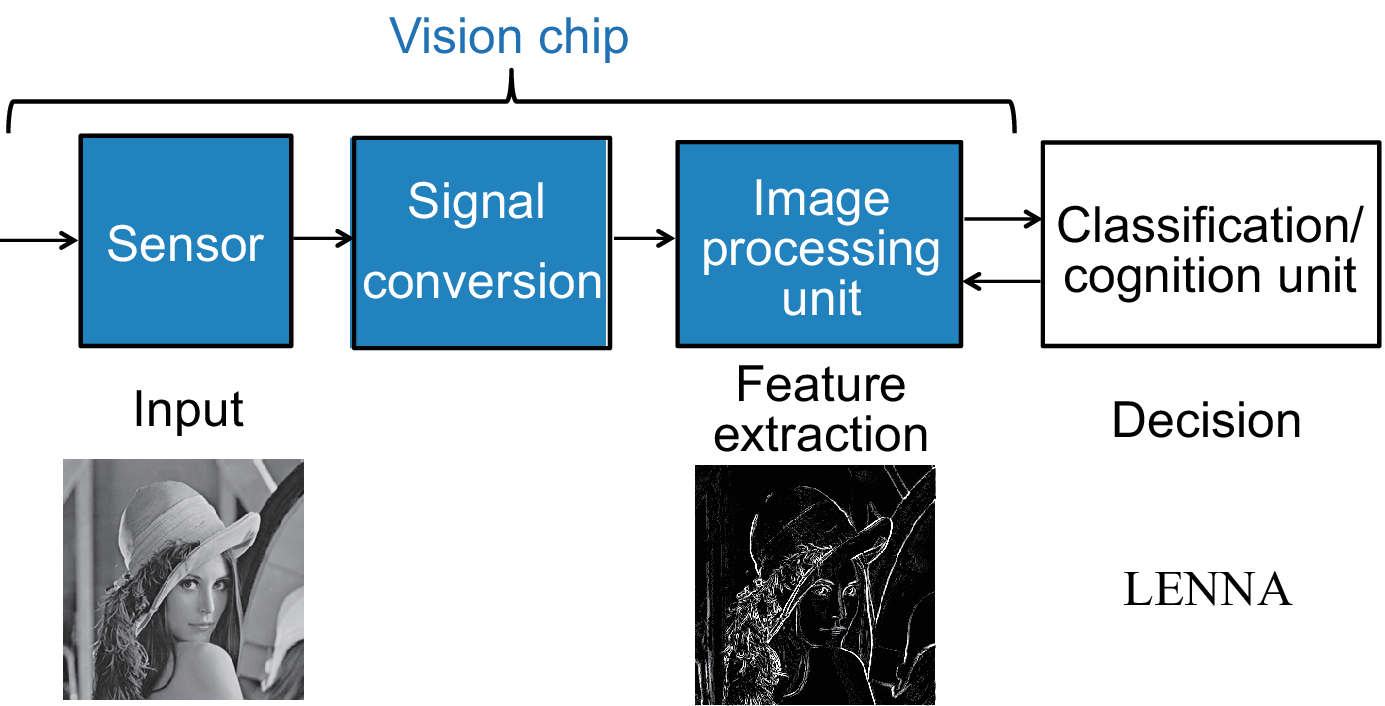}
\caption{Example of a vision chip used in a cognitive system.}
\label{vision_chip}
\vspace{0mm}
\end{center}
\end{figure}

This paper introduces an analog-to-stochastic converter using a magnetic tunnel junction (MTJ) device for massively parallel vision chips.
The vision chips are front-end image processors for feature extractions in cognitive computing as shown in Fig. \ref{vision_chip}.
The MTJ device \cite{MTJ} is often exploited as a non-volatile memory that stores one-bit information as a resistance and is used for MRAMs and content-addressable memories \cite{MRAM,CAM}.
The MTJ device is at one of two resistance states and the switching speed between the two states is the order of ns.
The switching behavior between the two states is probabilistic \cite{prob,prob_dev1,prob_dev2}.
Exploiting the probabilistic behavior, analog signals are directly converted to stochastic bit streams in the proposed analog-to-stochastic converter.
The signal conversion is theoretically described and is evaluated using device and circuit parameters to determine proper parameters for designing the analog-to-stochastic conversion.

The short version of this work was presented in \cite{NANOARCH}, but it described only the theoretical analysis of the analog-to-stochastic converter.
The extended work presented in this paper includes: (1) a more detailed theoretical analysis of the analog-to-stochastic converter by considering a voltage-bias effect, (2) considering a resistance variability of the MTJ device, (3) designing and simulating the analog-to-stochastic converter at the transistor level, and (4) simulating a simple vision chip including the proposed converter and a stochastic edge detector.

The rest of the paper is follows.
Section II reviews the MTJ device and describes the probabilistic behavior.
Section III  describes the stochastic computation. 
Section IV introduces the analog-to-stochastic converter using the MTJ device and shows the circuit implementation.
Section V describes a resistance-variability effect on the converter and a compensation technique of the variability.
Section VI evaluates the signal-conversion characteristic using device and circuit parameters and verifies the signal conversion using  a transistor-level simulator.
Section VII concludes this paper.

\section{Probabilistic Switching Behavior of MTJ Device}

\subsection {MTJ device}

\begin{figure}[t]
\begin{center}
\includegraphics[width=0.8\linewidth]{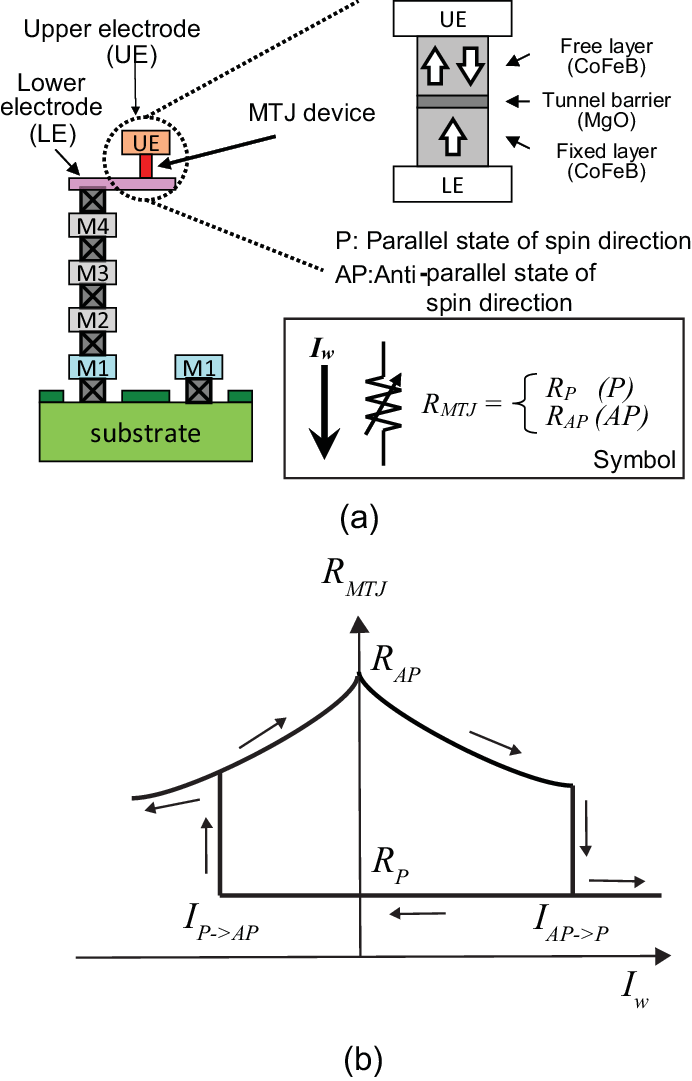}
\caption{Spin-transfer torque (STT) MTJ device: (a) stacked on a CMOS layer and (b) its R-I characteristic.}
\label{MTJ}
\vspace{0mm}
\end{center}
\end{figure}

Fig. \ref{MTJ} (a) shows a perpendicular spin-transfer torque (STT) MTJ device \cite{MTJ} stacked on a CMOS layer \cite{MTJ}.
The MTJ device consists of three main layers: a free layer, a tunnel-barrier layer, and a fixed layer, where there are a lot of additional layers in a real device.
A spin direction is fixed in the fixed layer while it can be changed to one of two directions in the free layer.
If the spin directions are the same in both layers, the MTJ device is on a parallel state that the MTJ resistance ($R_{MTJ}$) is low ($R_P$).
Otherwise, it is on an anti-parallel state that $R_{MTJ}$ is high ($R_{AP}$).

In the STT MTJ device, one-bit information is stored as a resistance using a current signal, $I_{w}$.
Fig. \ref{MTJ} (b) shows a characteristic between $R_{MTJ}$ and $I_{w}$.
The anti-parallel state of the MTJ device is changed to the parallel state using $I_{w}$ whose amount is $I_{AP->P}$ during a specific period, such as several ns (see \cite{MTJ}).
From the parallel to the anti-parallel state, the opposite direction of $I_{w}$ is used, where the amount of current is $I_{P->AP}$.

\subsection{Switching probability}

\begin{figure}[t]
\begin{center}
\includegraphics[width=0.6\linewidth]{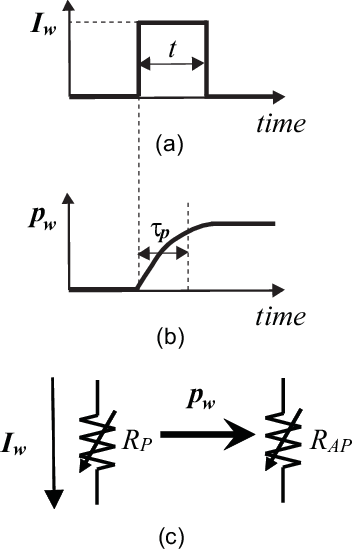}
\caption{Probabilistic behavior of MTJ device: (a) write current ($I_w$), (b) switching probability ($p_w$), and (c) switching behavior. The MTJ state is probabilistically changed to another state.}
\label{prob}
\vspace{0mm}
\end{center}
\end{figure}

The switching behavior of the MTJ device is probabilistic in \cite{prob,prob_dev1,prob_dev2}.
In several literatures, the probabilistic behaviour is exploited for MRAMs,  synapse systems, and random number generators  \cite{prob2,prob3,Spindice}.
Fig. \ref{prob} shows a probabilistic behavior of the MTJ device.
Suppose that the initial state of the MTJ device is parallel.
When one-bit information is written in the MTJ device, a write current signal, $I_{w}$, is applied during a specific period of $t$.
The switching probability of the MTJ device, $p_w$, is defined as follows \cite{prob}:
\begin{equation}
p_w=1-{\rm exp}(-t/\tau_p),
\label{eq:prob}
\end{equation}
where $\tau_p$ is the switching time constant.
Eq. (\ref{eq:prob}) is based on \cite{ISCAS} whose model derives from \cite{mtj_thermal}.
$\tau_p$ is calculated on either regions I or II described as follows.

To calculate $p_w$, a critical current, $I_c$, needs to be defined.
$I_c$ is differently defined in two regions: region I ($\tau_p$ $\gg$ $\tau_0$) and region II ($\tau_p$ $<$ several tens times $\tau_0$), where $\tau_0$ is the attempt time for thermal switching (around 1 ns) (see \cite{prob}).
In the region I (thermal activation region), the state of the MTJ device is mainly changed based on a thermal effect.
The thermal effect is expressed as $\ln(\tau_p/\tau_0)=(E/k_BT)$, where $E$ is the actual magnetic anisotropy energy of a magnetic cell, $k_B$ is the Boltzmann constant, and $T$ is the absolute temperature.
$I_{c}$ is defined as follows:
\begin{equation}
I_{c}=I_{c0t}(1-(k_BT/E)\ln(\tau_p/\tau_0)),
\label{eq:region1}
\end{equation}
where $I_{c0t}$ is the extrapolated $I_{c}$ at $\tau_p$ = $\tau_0$.
In the region II, the state of the MTJ device is mainly changed by a spin-injection effect.
$I_{c}$ is defined as follows:
\begin{equation}
I_{c}=I_{c0s}\Biggl(1+\frac{\tau_{relax}}{\tau_p}\ln\biggl(\frac{\pi/2}{\sqrt{k_BT/E}}\biggr)\Biggr),
\label{eq:region2}
\end{equation}
where $I_{c0s}$ is the extrapolated $I_{c}$ at $\tau_p$ $\sim$ $\infty$ and $\tau_{relax}$ is the relaxation time of a magnetic moment.
Using the probabilistic behavior, an analog-to-stochastic converter is designed for image processors based on stochastic computation.

\section{Stochastic computation}

\begin{figure}[t]
\begin{center}
\includegraphics[width=0.7\linewidth]{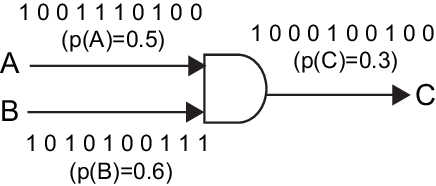}
\caption{Multiplier  based on stochastic computation designed using a 2-input AND gate.}
\label{stochastic}
\vspace{0mm}
\end{center}
\end{figure}

Stochastic computation was first introduced in 1960s \cite{stochastic_first} and has recently been exploited for some applications \cite{stochastic}, such as LDPC decoding \cite{ldpc1,ldpc2,ldpc3,ldpc4,ldpc5} and image processing \cite{image1,image2,gabor}.
Stochastic computation performs in probabilistic domain that the probability is represented by a random sequence of bits, called a {\it Bernoulli} sequence.
The probability is determined by the frequency of ones or zeros in the sequence that can be represented by many different sequences of bits.
For example, different sequences of bits (1010) and (0011) mean the same probability.

The advantage of stochastic computation is low-cost hardware implementation.
Fig. \ref{stochastic} shows a multiplier based on stochastic computation.
Let $p(A)$ = Pr(A=1) and $p(B)$ = Pr(B=1) be the probabilities represented by the two input bit streams and $p(C)$ = Pr(C=1) be the probability represented by the output bit stream.
The multiplication is described as $p(C)=p(A)*p(B)$ that is simply designed using a 2-input AND gate, where the computation performs bit-serially.
Hence, this technique results in efficient hardware implementations for algorithms that compute with probabilities such as LDPC decoders.

\begin{figure}[t]
\begin{center}
\includegraphics[width=1.0\linewidth]{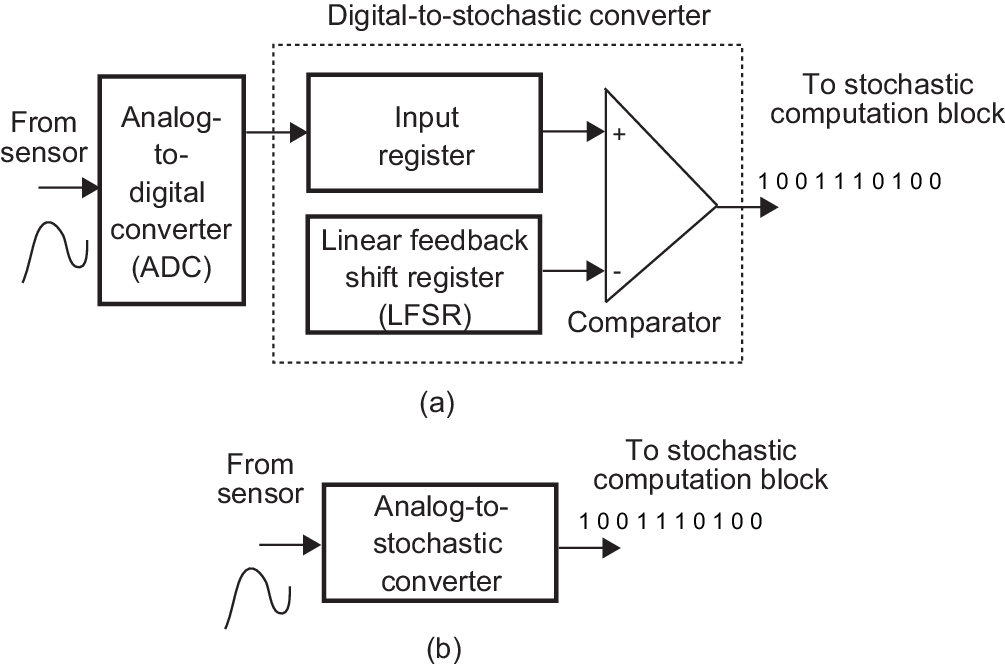}
\caption{Analog-to-stochastic converter: (a) conventional and (b) proposed. The proposed converter directly converts analog to stochastic signals to mitigate the signal-conversion overhead of the conventional system.}
\label{conversion}
\vspace{0mm}
\end{center}
\end{figure}

\begin{figure}[t]
\begin{center}
\includegraphics[width=1.0\linewidth]{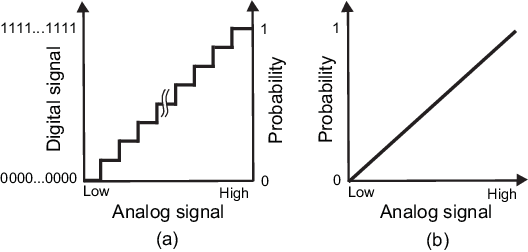}
\caption{Signal conversion: (a) conventional and (b) proposed.}
\label{signal}
\vspace{0mm}
\end{center}
\end{figure}

In contrast, the disadvantage of stochastic computation is a signal conversion.
Fig. \ref{conversion} (a) shows a conventional analog-to-stochastic converter for stochastic computation.
For example, in image processing, an analog signal is received from a sensor and is then converted to a digital signal using an analog-to-digital converter (ADC).
The digital signal is stored in registers and is then compared with random bits that typically generated using a linear feedback shift register (LFSR) to convert to a stochastic signal.
The relationship among the three signals is summarized in Fig. \ref{signal} (a).

The power dissipation of the ADC is a large portion of the total power dissipation in an image sensor (e.g. 65\% in  \cite{sensor}).
In addition, the digital-to-stochastic converter tends to be large in the stochastic circuits.
To estimate the overhead, an edge-detection circuit based on stochastic computation \cite{image1} is designed using Verilog-HDL and is synthesized using Synopsys Design Compiler on Silterra \SI{0.13}{\micro\meter} CMOS technology.
The synthesized result shows that the area of the digital-to-stochastic converter is 43.6\% of the total area when the bit width of the LFSRs is 10 bits.

To reduce the overhead of the signal conversion, an analog-to-stochastic converter is presented that the analog signal is directly converted to the stochastic signal as shown in Fig. \ref{conversion} (b).

\section{Analog-to-Stochastic Conversion Using MTJ Device}

\begin{figure}[t]
\begin{center}
\includegraphics[width=1.0\linewidth]{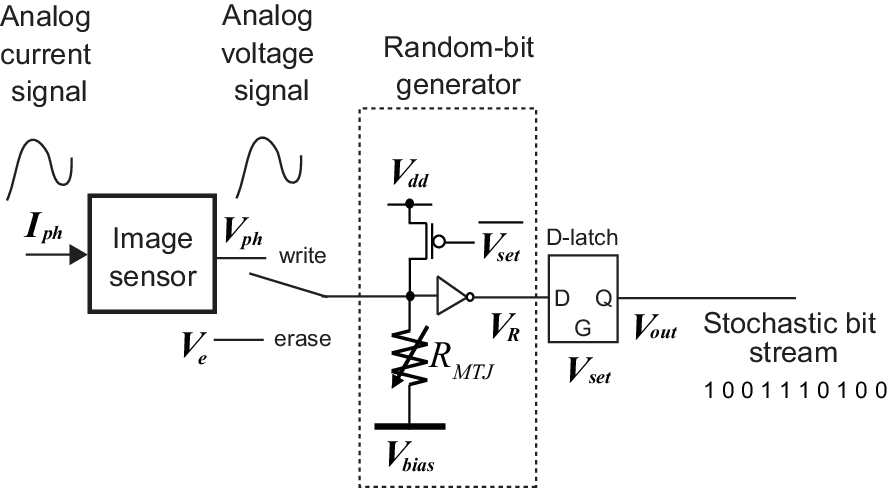}
\caption{Overall structure of the proposed analog-to-stochastic converter.}
\label{overall}
\vspace{0mm}
\end{center}
\end{figure}

\begin{figure}[t]
\begin{center}
\includegraphics[width=1.0\linewidth]{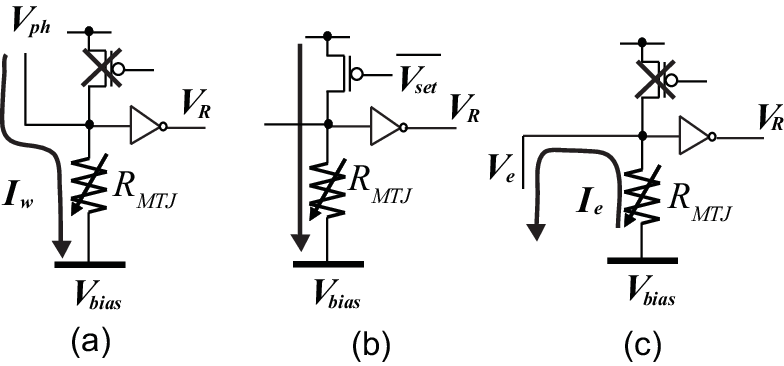}
\caption{Circuit of the analog-to-stochastic converter that operates at one of  : (a) write, (b) set, and (c) erase phases. One-bit information is stored at $p_w$ in the write phase and the stored MTJ state is set to the output in the set phase and finally the MTJ state is erased in the erase phase.}
\label{operation}
\vspace{0mm}
\end{center}
\end{figure}

\subsection{Overall structure}

Fig. \ref{overall} shows the proposed analog-to-stochastic converter using the MTJ device.
Suppose that an analog current signal is received in a logarithm image sensor that realizes a high dynamic range  \cite{log_sensor1,log_sensor2,log_sensor3}.
The analog voltage signal is the output of the image sensor and is the input of the analog-to-stochastic converter.
The analog-to-stochastic converter is designed using three transistors and one MTJ device.
The output of the converter connects to a latch that stores a stochastic bit generated.

Fig. \ref{operation} shows the circuit operation of the proposed analog-to-stochastic converter.
The converter operates at one of three phases: write, set, and erase.
After the end of the erase phase, the phase is back to the write phase.

\subsection{Write phase}

In the write phase, the analog input current signal, $I_{ph}$, is converted to the analog output voltage signal, $V_{ph}$, as follows:
\begin{equation}
V_{ph} = V_{dd}-n\frac{k_BT}{q}\ln\Bigl(\frac{I_{ph}}{I_{d0}}\Bigr),
\label{eq:sensor}
\end{equation}
where $n$ is 1.3 - 2, $I_{d0}$ is the process dependent parameter and $q$ is the elementary charge \cite{log_sensor2}.
Then, $V_{ph}$ is applied to the analog-to-stochastic converter to generate a write current signal, $I_{w}$, defined as follows:
\begin{equation}
I_{w}= \frac{V_{dd}-V_{bias}}{R_{MTJ}}-n\frac{k_BT}{qR_{MTJ}}\ln\Bigl(\frac{I_{ph}}{I_{d0}}\Bigr),
\label{eq:current}
\end{equation}
where $V_{bias}$ is the bias voltage to control $I_{w}$.

The resistance, $R_{MTJ}$, varies depending on a bias voltage to the MTJ device, $V_b$(=$V_{ph}$-$V_{bias}$) when a current flows through the MTJ device \cite{MTJ}.
Especially, the resistance at the anti-parallel state is more sensitive to the bias voltage than that at the parallel state.
To mitigate the bias-voltage effect, the parallel state is used as the initial state of the MTJ device in the converter.
The resistance at the parallel state with the bias voltage effect is defined as $R_{Pb}$ that is described as:
\begin{equation}
R_{Pb}=R_{P}(1+BC1|V_b|+BC2|V_b|^2),
\label{eq:bias}
\end{equation}
where BC1 and BC2 are the first and second-bias coefficients, respectively.
$R_{P}$ is the resistance at the parallel state when the bias voltage is 0.
$I_{w}$ is redefined as 
\begin{equation}
I_{w}= \frac{V_{dd}-V_{bias}}{R_{Pb}}-n\frac{k_BT}{qR_{Pb}}\ln\Bigl(\frac{I_{ph}}{I_{d0}}\Bigr).
\label{eq:current_bias}
\end{equation}

Suppose that the switching characteristic of the MTJ device is determined by the region II described in equation (\ref{eq:region2}).
The switching time constant, $\tau_p$, is defined as follows:
\begin{equation}
\tau_p=\frac{I_{c0s}\tau_{relax}\ln\frac{\pi/2}{\sqrt{k_BT/E}}}{{I_{w}-I_{c0s}}}.
\label{eq:taup}
\end{equation}
A non-switching probability, $\overline{p_{w}}$, is defined as $1-p_{w}$,  where $p_w$ is described in equation (\ref{eq:prob}).
The non-switching probability in logarithm domain, $\ln\overline{p_w}$, is described as follows:
\begin{equation}
\ln\overline{p_w}=-t/\tau_p.
\label{eq:log_prob}
\end{equation}
Using equations (\ref{eq:taup}) and (\ref{eq:log_prob}), $\ln\overline{p_w}$ is defined as follows:
\begin{equation}
\ln\overline{p_w}=-\alpha(I_{w}-I_{c0s}),
\label{eq:log_prob2}
\end{equation}
where $\alpha$ is $t/\Bigl(I_{c0s}t_{relax}\ln\Bigl(\frac{\pi/2}{\sqrt{k_BT/E}}\Bigr)\Bigr)$.
Using equations (\ref{eq:current_bias}) and (\ref{eq:log_prob2}),  $\ln\overline{p_w}$ is redefined as follows:
\begin{equation}
\ln\overline{p_w}=\beta\ln\Bigl(\frac{I_{ph}}{I_{d0}}\Bigr)+\alpha\Bigl(I_{c0s}-\frac{(V_{dd}-V_{bias})}{R_{Pb}}\Bigr),
\label{eq:log_prob3}
\end{equation}
where $\beta$ = $\alpha n\frac{k_BT}{qR_{Pb}}$.
$\beta$ has to be equal to 1 to ensure a linear relationship between the analog input signal, $I_{ph}$, and the probability of the  stochastic bit stream, $p_w$.
The linear relationship is required for the analog-to-stochastic converter.
From this, one derives $\alpha$ as follows:
\begin{equation}
\alpha = \frac{qR_{Pb}}{nk_BT},
\label{eq:alpha}
\end{equation}
 and thus the pulse duration, $t$,  is given by:
\begin{equation}
t = \Bigl(I_{c0s}t_{relax}\ln\Bigl(\frac{\pi/2}{\sqrt{k_BT/E}}\Bigr)\Bigr) \cdot \frac{qR_{Pb}}{nk_BT}.
\label{eq:time}
\end{equation}

\subsection{Set and erase phases}

Once one-bit information is stored in the MTJ device at $p_w$, the converter operates at the set phase shown in Fig. \ref{operation} (b).
At the set phase, $V_{set}$ is high and hence the voltage signal, $V_{R}$, is determined by the MTJ state defined as follows:
\begin{equation}
V_{R}=\left\{ 
\begin{array}{ll}
{\rm High \ (``1")} & \mbox{if $R_{MTJ}$=$R_{P}$}\\
{\rm Low \ (``0")} & \mbox{otherwise} \\
\end{array} \right.
\label{eq:output}
\end{equation}
Then, $V_{R}$ is stored in the latch next to the converter as shown in Fig. \ref{overall}.

After outputting the stochastic bit, the converter operates at the erase phase shown in Fig. \ref{operation} (c).
At the erase phase, the erase current, $I_{e}$, is applied to change $R_{MTJ}$ back to $R_{P}$, where the current direction of $I_{e}$, is opposite to that of $I_{w}$.
Finally, the operation phase returns to the write phase to generate a new stochastic bit.

\section{Consideration of MTJ Variability}

\subsection{Theoretical analysis}

\begin{figure}[t]
\begin{center}
\includegraphics[width=1.0\linewidth]{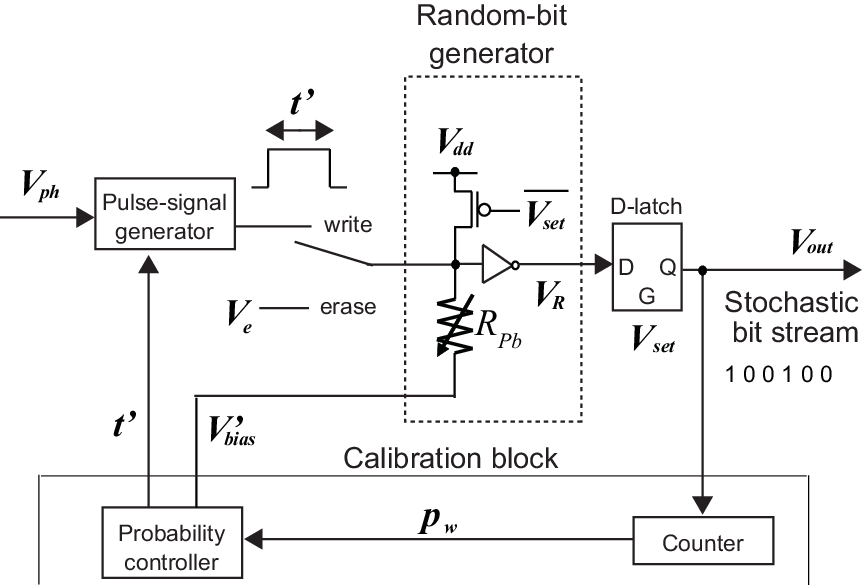}
\caption{Overall structure of the proposed analog-to-stochastic converter considering a resistance variability of the MTJ device. $t'$ and $V'_{bias}$ are controlled to compensate the variability as a calibration step.}
\label{overall2}
\vspace{0mm}
\end{center}
\end{figure}

In this section, a resistance variability of an MTJ device is considered.
In the analog-to-stochastic converter, the resistance variability affects the write current in equation (\ref{eq:current_bias}), and hence the switching probability in equation (\ref{eq:log_prob3}).
In this case, an analog input signal is not properly converted to a stochastic bit stream in the converter.

Here, the resistance variability is defined by $\Delta R$.
The write current considering the resistance variability, $I_w'$, is defined using equation (\ref{eq:current_bias}) as follows:
\begin{equation}
I_{w}' \simeq \frac{V_{dd}-V'_{bias}}{(R_{Pb}+\Delta R)}-n\frac{k_BT}{q(R_{Pb}+\Delta R)}\ln\Bigl(\frac{I_{ph}}{I_{d0}}\Bigr),
\label{eq:current_bias2}
\end{equation}
where  $V_{bias}'$ is an updated parameter of $V_{bias}$ in case of considering the MTJ variability.
Suppose that $R_{Pb}$ defined in equation (\ref{eq:bias}) is not affected by changing $V_{bias}$ to $V'_{bias}$.

In addition, the switching probability considering the MTJ variability, $p_w'$, is defined using equation (\ref{eq:log_prob3}) as follows:
\begin{equation}
\ln\overline{p_w'} \simeq \beta'\ln\Bigl(\frac{I_{ph}}{I_{d0}}\Bigr)+\alpha'\Bigl(I_{c0s}-\frac{(V_{dd}-V_{bias}')}{(R_{Pb}+\Delta R)}\Bigr),
\label{eq:log_prob4}
\end{equation}
where $\alpha'$ and  $\beta'$ are updated parameters of $\alpha$ and $\beta$, respectively, in case of considering the MTJ variability.
As shown in the two equations, the write current and the switching probability are affected by $\Delta R$.
To compensate the $\Delta R$ effect, three parameters, such as $\alpha'$, $\beta'$, and $V_{bias}$ are controlled.

First, let us consider $\beta'$ in equation (\ref{eq:log_prob4}).
As $\beta$ is $\alpha n\frac{k_BT}{qR_{Pb}}$, $\beta'$ is defined as follows:
\begin{equation}
\beta' \simeq \frac{t'}{\Bigl(I_{c0s}t_{relax}\ln\Bigl(\frac{\pi/2}{\sqrt{k_BT/E}}\Bigr)\Bigr)} \cdot \frac{nk_BT}{q(R_{Pb}+\Delta R)},
\label{eq:beta}
\end{equation}
where $t'$ is an updated parameter of $t$ in case of  considering the resistance variability.
In equation (\ref{eq:beta}), there is a variable parameter, $t'$, to compensate the resistance variability.
To realize the linear relationship between the analog input signal, $I_{ph}$, and the probability of the  stochastic bit stream, $p_w$ under the resistance variability, $\beta'$ has to be 1.
Hence, the resistance variability in terms of $\beta'$ is compensated by the write attempt time, $t'$, as follows:
\begin{equation}
t' \simeq \Bigl(I_{c0s}t_{relax}\ln\Bigl(\frac{\pi/2}{\sqrt{k_BT/E}}\Bigr)\Bigr) \cdot \frac{q(R_{Pb}+\Delta R)}{nk_BT}.
\label{eq:time2}
\end{equation}
Using equations (\ref{eq:time}) and (\ref{eq:time2}), $t'$ is redefined as follows:
\begin{equation}
t' \simeq t(1+\frac{\Delta R}{R_{Pb}}).
\label{eq:time3}
\end{equation}
When $\beta'$ is equal to 1, $\alpha'$ based on equation (\ref{eq:alpha}) is given by:
\begin{equation}
\alpha' \simeq \frac{q(R_{Pb}+\Delta R)}{nk_BT}.
\label{eq:alpha2}
\end{equation}

Next, let us consider $\alpha'$ and $\frac{(V_{dd}-V_{bias}')}{(R_{Pb}+\Delta R)}$ in equation (\ref{eq:log_prob4}).
To maintain the same switching probability under the resistance variability, we need to satisfy an equation using equations (\ref{eq:log_prob3}) and (\ref{eq:log_prob4}) as follows:
\begin{equation}
\alpha\Bigl(I_{c0s}-\frac{(V_{dd}-V_{bias})}{R_{Pb}}\Bigr) \simeq \alpha'\Bigl(I_{c0s}-\frac{(V_{dd}-V_{bias}')}{(R_{Pb}+\Delta R)}\Bigr)
\label{eq:satisfy}
\end{equation}
Using (\ref{eq:alpha}), (\ref{eq:alpha2}), and (\ref{eq:satisfy}), $V_{bias}'$ is defined as follows:
\begin{equation}
V_{bias}' \simeq V_{bias} - \Delta R I_{c0s}.
\label{eq:vbias}
\end{equation}
Based on the theoretical analysis, we can control $t'$ and $V_{bias}$ to maintain the same switching probability under the resistance variability.

\subsection{Compensation of resistance variability}

Fig. \ref{overall2} shows a block diagram of the proposed analog-to-stochastic converter considering the resistance variability.
It consists of a pulse-signal generator, a random bit generator, a counter, and a probability controller.
Suppose that the two parameters, $t'$ and $V_{bias}'$, are set as a calibration step before using the converter.

First, an input-voltage signal, $V_{ph}$, is tested using default parameters of $t$ and $V_{bias}$ to generate a stochastic bit stream whose probability is 50\%.
Then, the number of 1's in the bit stream is counted and the probability is checked in the probability controller.
If the probability is not 50\%, then $V_{bias}'$ is controlled.
For example, if the probability is less than 50\%, the MTJ resistance is larger than that without variability.

Once a probability of 50\% is obtained, different input-voltage signals are tested to generate different probabilities.
If the relationship between the analog input signal and the probability is not linear, $t'$ is controlled to set $\beta'$ 1.
After setting the two parameters desired, the analog-to-stochastic converter is ready to start.

\section{Evaluation}

\subsection{Theoretical analysis}

\begin{table}[t]
\caption{Device and circuit parameters.}
\label{parameter}
\centering
\begin{tabular}{c|c|c}
\hline
Parameters & Name & Value \\
\hline
\hline
$V_{dd}$ & Supply voltage & 1.2 V \\
$I_{d0}$ & - & 0.1 nA \\
$n$ & - & 2 \\
$T$ & Absolute temperature & 300 K \\
\hline
$I_{c0s}$ & Critical current & \SIrange{50}{200}{\micro\ampere} \\
$R_{P}$ & Resistance (P) & 1 k$\Omega$ \\
$R_{AP}$ & Resistance (AP) & 3 k$\Omega$ \\
$E/k_BT$ & - & 60 \\
$t_{relax}$  & Relaxation time & 500 ps \\
\hline
t & Attempt time of $I_w$ & 1 - 10 ns \\
$V_{bias}$ & Bias voltage & 0 - 0.4 V\\
\hline
\end{tabular}
\end{table}

The proposed analog-to-stochastic converter is evaluated using device and circuit parameters shown in Table \ref{parameter}.
Parameters, $I_{d0}$ and $n$,  in image sensors are derived from \cite{log_sensor2} and device parameters, such as $I_{c0s}$, $R_{P}$, $R_{AP}$, $E/K_BT$, and $t_{relax}$, are derived from \cite{MTJ,prob}.
The circuit parameters are based on ASPLA 90nm CMOS technology and the MTJ-device parameters are based on a 100nm MTJ technology.

\begin{figure}[t]
\begin{center}
\includegraphics[width=1.0\linewidth]{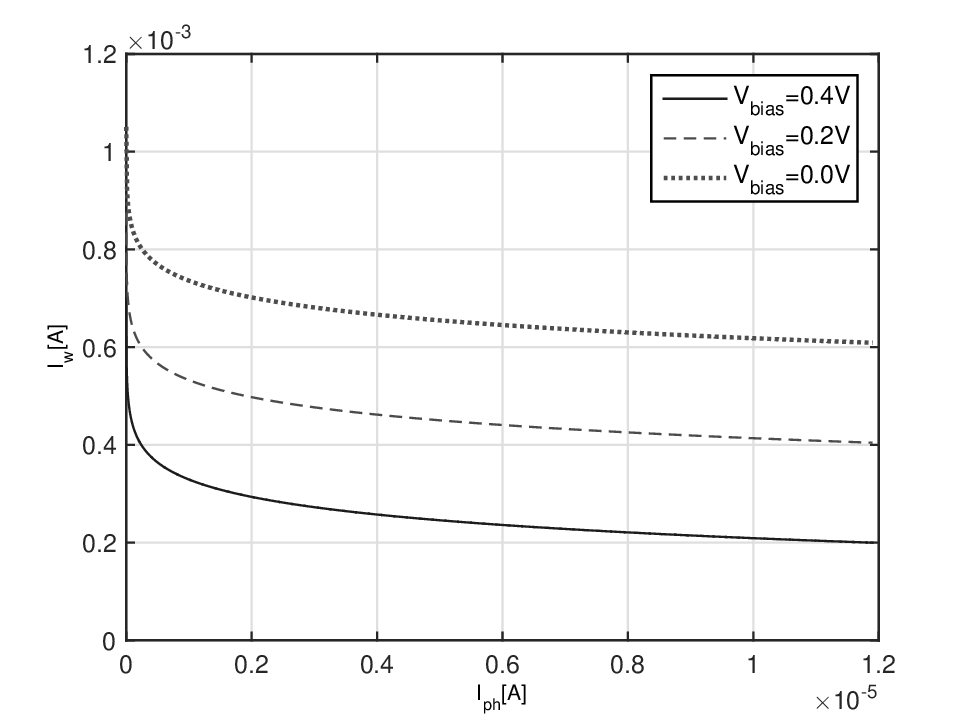}
\caption{Relationship between the write current at the MTJ device ($I_w$) and the analog input current ($I_{ph}$).}
\label{graph1}
\vspace{0mm}
\end{center}
\end{figure}

Fig. \ref{graph1} shows a relationship between the write current at the MTJ device, $I_w$, and the analog input current, $I_{ph}$, based on equation (\ref{eq:current_bias}).
In equation (\ref{eq:current_bias}), $I_{ph}$ is only a variable.
The other parameters, except $V_{bias}$ are device dependent and constant.
$V_{bias}$ are used to control the amount of $I_w$.

\begin{figure}[t]
\begin{center}
\includegraphics[width=1.0\linewidth]{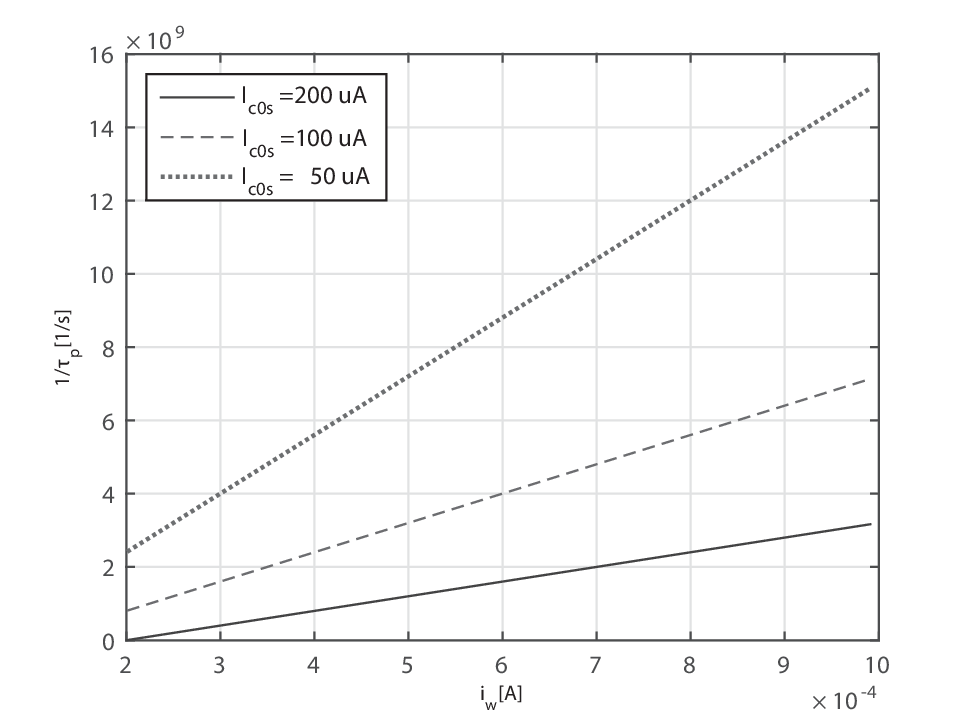}
\caption{Relationship between inverse of the switching time constant ($\tau_p$) and $I_w$ when $V_{bias}$ is 0.4 V.}
\label{graph2}
\vspace{0mm}
\end{center}
\end{figure}

Fig. \ref{graph2} shows a relationship between the inverse of the switching time constant, $\tau_p$, and $I_w$ based on equation (\ref{eq:taup}) when $V_{bias}$ is 0.4 V.
When the critical current, $I_{c0s}$, is decreased, $\tau_p$ is decreased.
It means that the MTJ device is more easily switched to another state using the same amount of current, $I_w$, when $I_{c0s}$ is smaller.
$I_{c0s}$ is decreased when the MTJ technology is advanced \cite{MTJ11nm}, reducing the write energy and the write attempt time, $t$.
%


\begin{figure}[t]
\begin{center}
\includegraphics[width=1.0\linewidth]{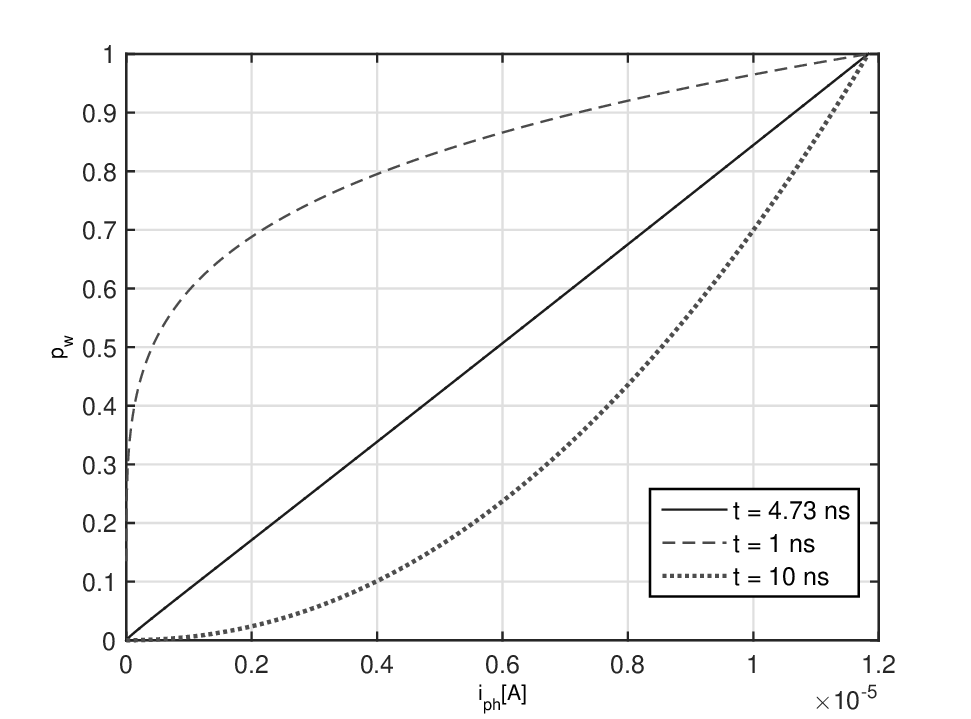}
\caption{Relationship between the non-switching probability of the MTJ device  ($\overline{p_w}$) and $I_{ph}$ when $V_{bias}$ is 0.4 V and $I_{c0s}$ is \SI{200}{\micro\ampere}. The analog-to-stochastic converter is properly designed by setting $t$ = 4.73 ns as the linear relationship is realized.}
\label{graph4}
\vspace{0mm}
\end{center}
\end{figure}

Fig. \ref{graph4} shows a relationship between the non-switching probability of the MTJ device, $\overline{p_w}$, and $I_{ph}$, when $V_{bias}$ is 0.4 V and $I_{c0s}$ is \SI{200}{\micro\ampere} based on equation (\ref{eq:log_prob3}).
The attempt time, $t$, varies from 1 to 10 ns.
When $I_{ph}$ is increased, $I_w$ is decreased, reducing $p_w$.
Depending on $t$, the parameter, $\beta$, in equation (\ref{eq:log_prob3}) is changed, affecting the slope in Fig. \ref{graph4}.
When $\beta$ is set to 1 by choosing $t$ = 4.73 ns, the relationship between $p_w$ and $I_{ph}$ is linear, realizing the analog-to-stochastic conversion.

\subsection{Theoretical analysis with considering MTJ variability}

\begin{figure}[t]
\begin{center}
\includegraphics[width=1.0\linewidth]{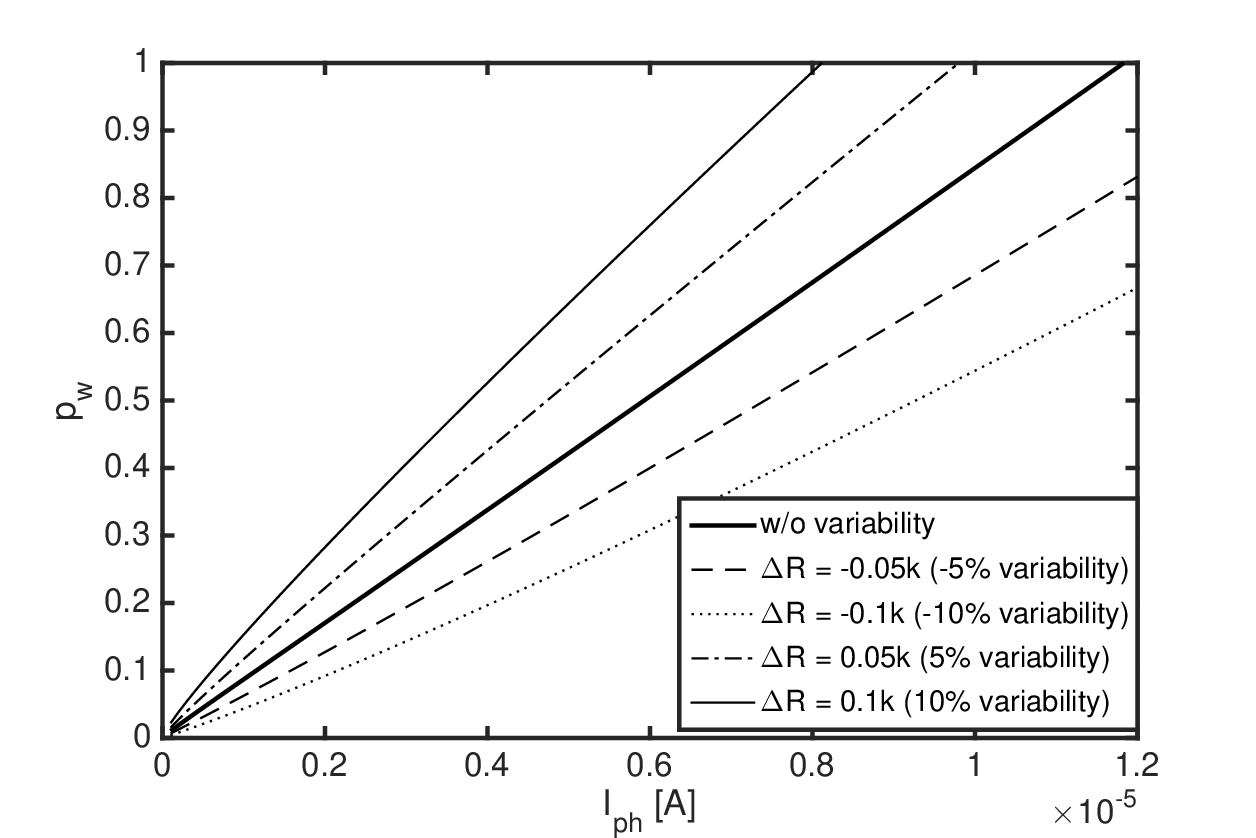}
\caption{Relationship between the non-switching probability of the MTJ device  $\overline{p_w'}$ and $I_{ph}$ under the resistance variability of the MTJ device, where  $V_{bias}$ is 0.4 V, $I_{c0s}$ is \SI{200}{\micro\ampere} and $t$ is 4.73 ns.}
\label{graph11}
\vspace{0mm}
\end{center}
\end{figure}

Fig. \ref{graph11} shows a relationship between $\overline{p_w'}$ and $I_{ph}$ under the resistance variability of the MTJ device, where $V'_{bias}$ is 0.4 V, $I_{c0s}$ is \SI{200}{\micro\ampere} and $t$ is 4.73 ns.
The relationship is given by equation (\ref{eq:log_prob4}).
A resistance variability of 10 \% at most is considered.
Compared with the relationship without the variability, the switching probabilities considering the variability are different.

\begin{figure}[t]
\begin{center}
\includegraphics[width=1.0\linewidth]{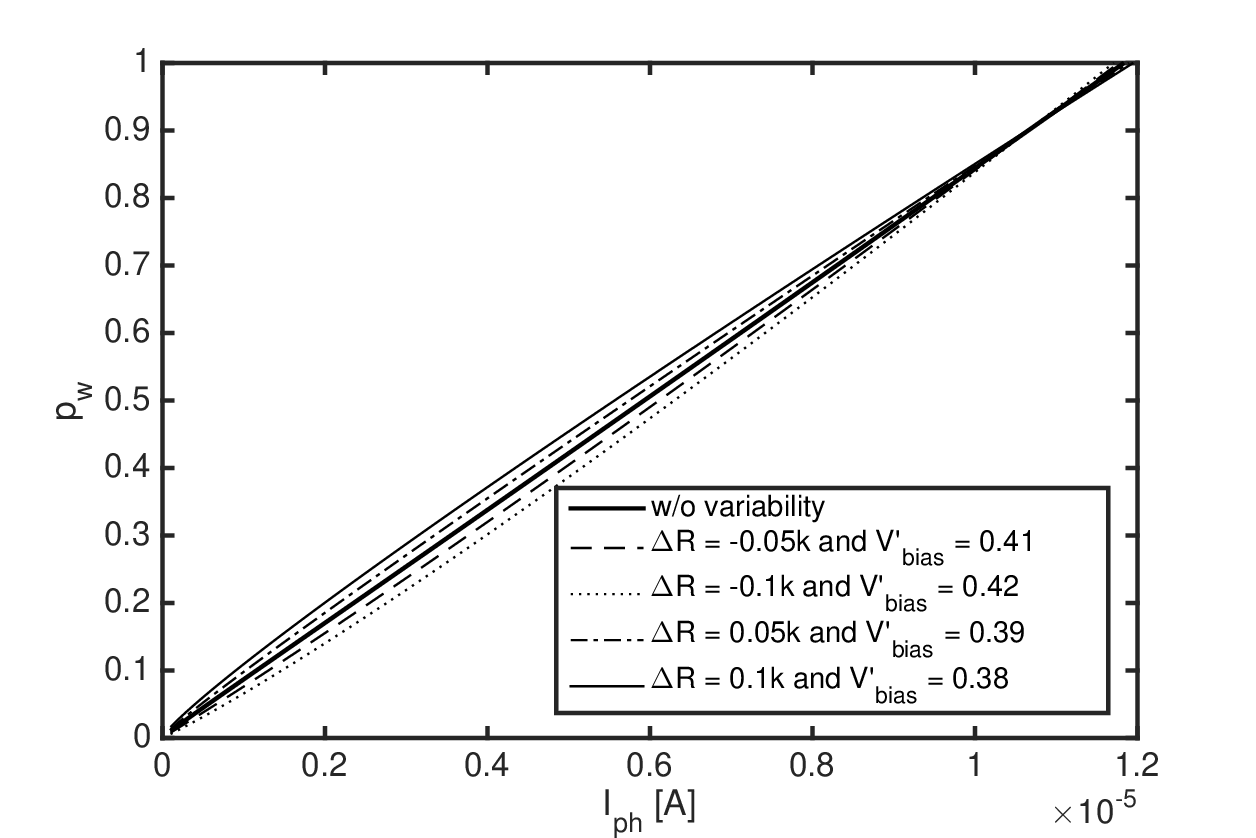}
\caption{Relationship between the non-switching probability of the MTJ device  $\overline{p_w'}$ and $I_{ph}$ under the resistance variability, where $I_{c0s}$ is \SI{200}{\micro\ampere} and $t$ is 4.73 ns. The resistance variability is compensated by controlling only $V'_{bias}$.}
\label{graph12}
\vspace{0mm}
\end{center}
\end{figure}

To compensate the variability, only $V'_{bias}$ is controlled as shown in Fig. \ref{graph12}.
$V'_{bias}$ are determined based on equation (\ref{eq:vbias}).
After setting $V'_{bias}$, the switching probabilities are close to that without the resistance variability.
However, the linear relationships between the switching probabilities and the analog input current signal are not obtained.

\begin{figure}[t]
\begin{center}
\includegraphics[width=1.0\linewidth]{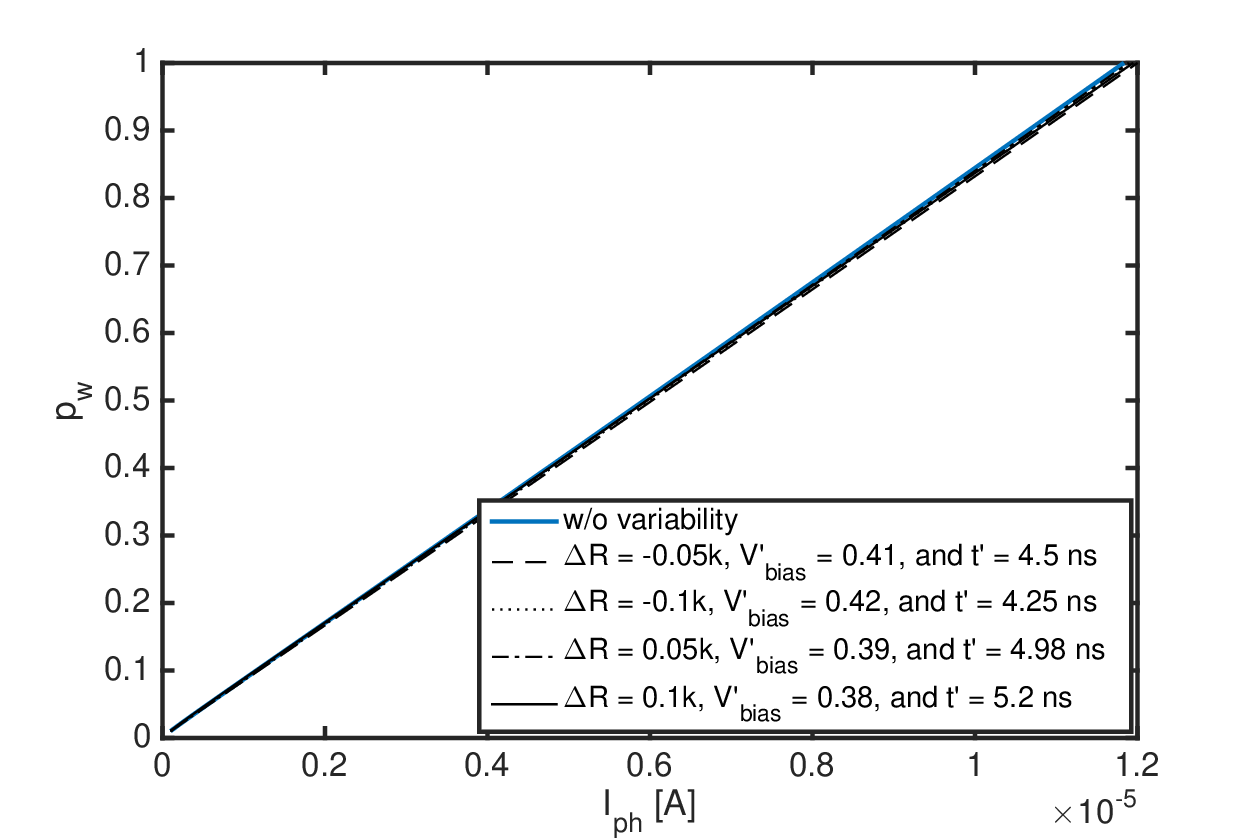}
\caption{Relationship between the non-switching probability of the MTJ device  $\overline{p_w}$ and $I_{ph}$ under the resistance variability where  $I_{c0s}$ is \SI{200}{\micro\ampere}. The resistance variability is compensated by controlling  $V'_{bias}$ and $t'$.}
\label{graph13}
\vspace{0mm}
\end{center}
\end{figure}

To control both $V_{bias'}$ and $t'$, the relationships between the switching probabilities and the analog input current signal are almost linear under the resistance variability as shown in Fig. \ref{graph13}.
$t'$ are determined based on equation (\ref{eq:time3}).

\subsection{Simulation results}

\begin{figure}[t]
\begin{center}
\includegraphics[width=1.0\linewidth]{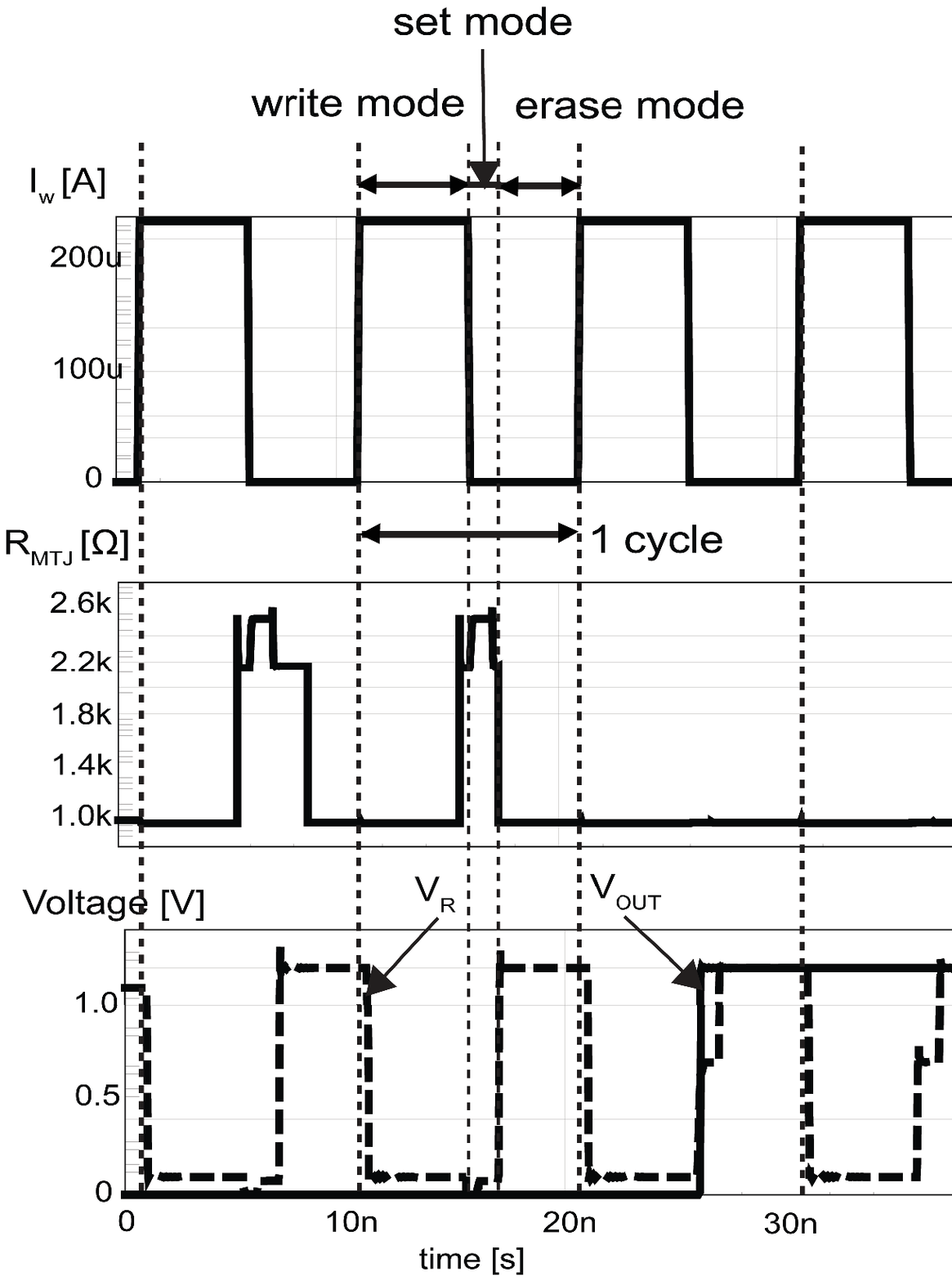}
\caption{Simulated waveforms of the proposed analog-to-stochastic converter in 90nm CMOS and 100nm MTJ technologies using NS-SPICE. The write current, $I_w$ is \SI{236}{\micro\ampere} corresponding to the switching probability, $p_w$, of 50\%.}
\label{waveform1}
\vspace{0mm}
\end{center}
\end{figure}

To verify the circuit operation of the proposed analog-to-stochastic converter,  the proposed converter is designed using device and circuit parameters used in Fig.  \ref{graph4}.
Fig. \ref{waveform1} shows simulated waveforms of the proposed analog-to-stochastic converter.
It is simulated using NS-SPICE \cite{ISCAS} in 90nm CMOS and 100nm MTJ technologies.
NS-SPICE is a transistor-level simulator that can handle both the transistors and the MTJ devices.
The cycle time of the converter is set to 10 ns, where the write phase is 5 ns, and the set phase is 1 ns, and the erase phase is 4 ns.
At the write phase, there is a write current during 4.73 ns and no current during 0.27 ns.
The write current, $I_w$, is \SI{236}{\micro\ampere} corresponding to the switching probability, $p_w$, of 50\%.

In this simulation, the proposed converter generates three stochastic bits.
Note that the resistance of the MTJ device is not the same as $R_{P}$ or $R_{AP}$ because of the bias voltage effect.
At the first and the second trials, the resistance of the MTJ device is changed from the parallel to the anti-parallel state at the write phase.
Hence, the output of the converter, $V_{OUT}$ is ``0".
In contrast, at the third trial, the resistance of the MTJ device is not changed even if $I_w$ is applied to the MTJ device.
In this case, $V_{OUT}$ is ``1".

\begin{figure}[t]
\begin{center}
\includegraphics[width=0.9\linewidth]{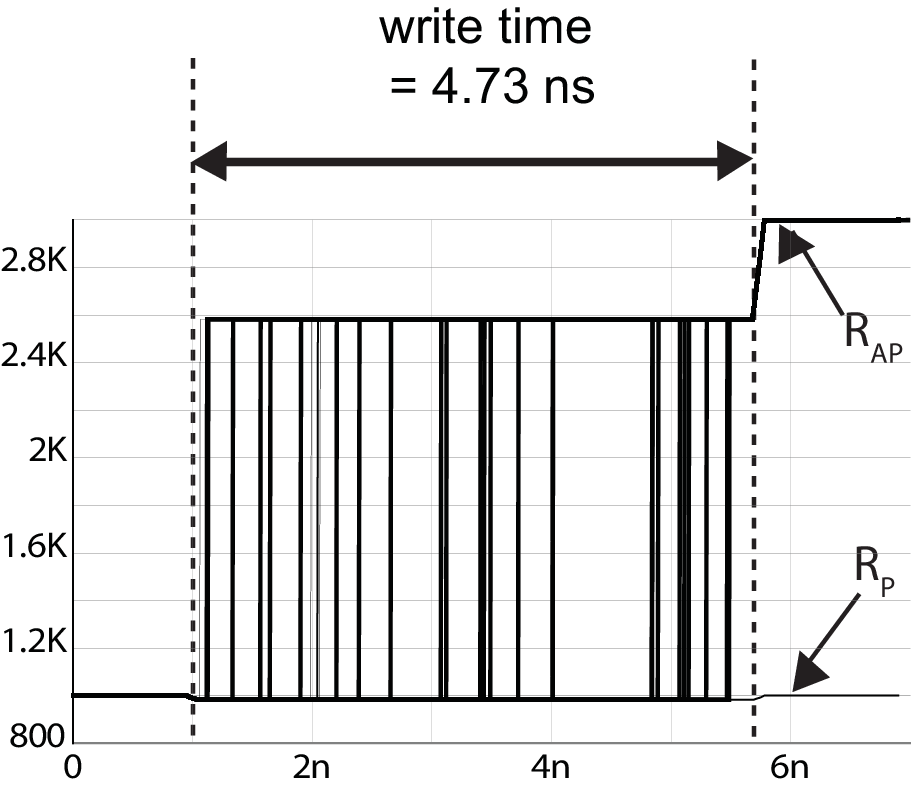}
\caption{Monte-carlo simulation of the proposed analog-to-stochastic converter, where $I_w$ is \SI{236}{\micro\ampere} corresponding to $p_W$=50\% and the number of trials is 100.}
\label{waveform2}
\vspace{0mm}
\end{center}
\end{figure}

Fig. \ref{waveform2} shows a monte-carlo simulation of the proposed analog-to-stochastic converter at the write phase.
The number of trials is 100 and $I_w$ is \SI{236}{\micro\ampere} corresponding to $p_w$ of 50\%.
The simulation waveforms show that the switching behavior of the MTJ device is probabilistic and the switching timing is random.
In this simulation, the resistance of the MTJ device is changed to $R_{AP}$ of 3 k$\Omega$ at 50\% after writing a bit to the MTJ device.


%
%

\subsection{Design example of a vision chip for cognitive systems}

\begin{figure}[t]
\begin{center}
\includegraphics[width=1.0\linewidth]{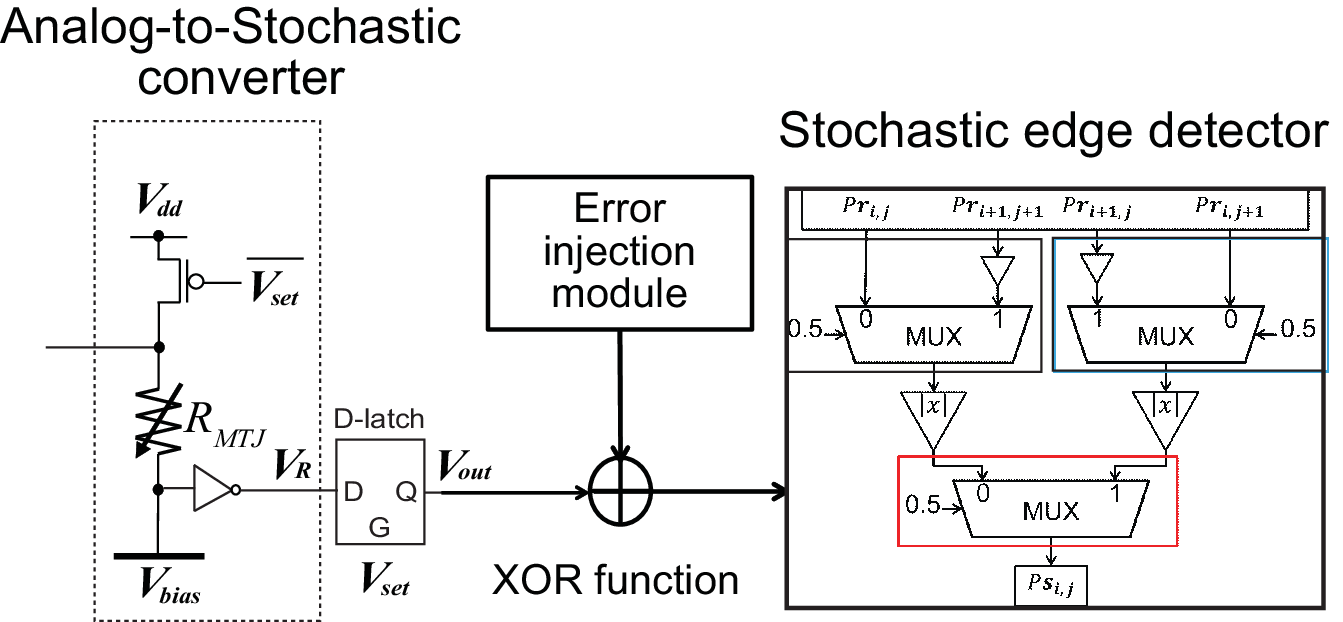}
\caption{Simulation model of a simple vision chip including an error-injection module in MATLAB.}
\label{edge_detection}
\vspace{0mm}
\end{center}
\end{figure}

Fig. \ref{edge_detection} shows a simulation model of a simple vision including an error-injection module.
Suppose the simple vision chip receives analog input signals of an image from a sensor and extracts the feature of the image for cognitive systems.
The feature extraction unit is designed using a stochastic edge detector \cite{image1,image2}.
The simulation model is designed and evaluated using MATLAB, where the analog-to-stochastic converter is simply modelled as a random bit generator.
It is because the precise model of the MTJ device requires transistor-level modelling.
It is also possible to include NS-SPICE in MATLAB to emulate the switching behavior of the MTJ device in \cite{cosimulation}.
The simulation model also includes an error injection module that models wrong bits that might be generated in the analog-to-stochastic converter.

\begin{figure}[t]
\begin{center}
\includegraphics[width=1.0\linewidth]{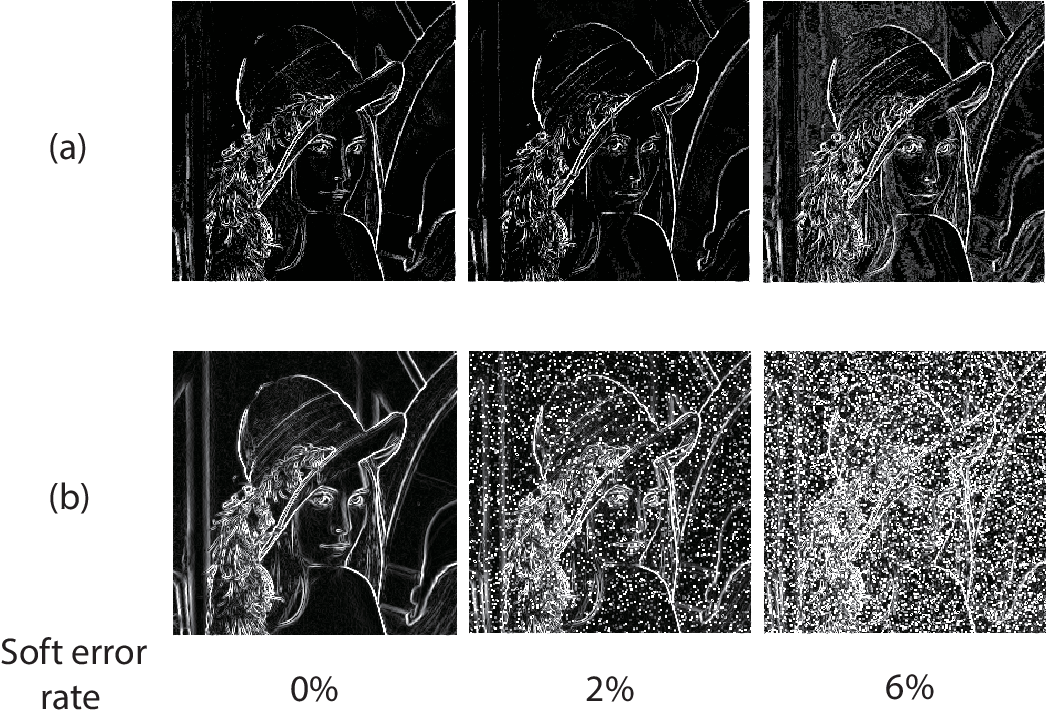}
\caption{Edge-detection results under input soft errors (a) stochastic implementation and  (b) conventional binary implementation.}
\label{LENNA}
\vspace{0mm}
\end{center}
\end{figure}

Fig. \ref{LENNA} (a) shows stochastic edge-detection results based on Fig. \ref{edge_detection} under input soft errors.
A standard grayscale test image, LENNA, is used to evaluate the error effects.
The number of stochastic bits is set to 1,000.
The simulated images show that the image qualities are not strongly affected by input errors, because stochastic computation is highly robust against soft errors.
In contrast, the conventional binary implementation heavily suffers from soft errors as shown in Fig. \ref{LENNA} (b).
As a result, the stochastic edge-detection result would be useful for cognitive systems, even if the proposed converter generates wrong stochastic bits at a small error rate.

\section{Conclusion}

In this paper, the analog-to-stochastic converter using MTJ devices has been proposed for the stochastic computation based vision chips.
Thanks to the probabilistic behavior of MTJ devices, analog signals from image sensors are directly and area-efficiently converted to stochastic bit streams without power-and-area hungry analog-to-digital and digital-to-stochastic converters.
The MTJ based  analog-to-stochastic conversion is theoretically described and is evaluated using device and circuit parameters.
As a result, a linear relationship between the analog signal and the probability of stochastic bit stream is realized by setting a specific attempt time of the write current.
In addition, the resistance variability of the MTJ device is theoretically analyzed.
It shows that the variability can be compensated by controlling the attempt time of writing and a bias voltage at a calibration step before using the converter.
Based on the theoretical analysis, the analog-to-stochastic converter is designed using  in 90nm CMOS and 100nm MTJ technologies.
The proposed converter simulated using NS-SPICE generates a stochastic bit stream.
In addition, the simple vision chip including an analog-to-stochastic converter and a stochastic edge detector is designed and simulated in MATLAB for future cognitive systems.

Future work includes a more detailed circuit implementation of the proposed converter including the calibration block for fabrication.
It also includes designing a whole cognitive system including the stochastic vision chip based on the analog-to-stochastic converter.

\section*{Acknowledgment}
This work was supported by JSPS KAKENHI Grant Number 26700003 and MEXT Brainware LSI Project.
This work is supported by VLSI Design and Education Center (VDEC), The University of Tokyo with the collaboration with Synopsys Corporation.
%



\end{document}